\documentclass[letter]{aa}

\usepackage{graphicx}
\usepackage{txfonts}
\usepackage{hyperref}
\usepackage{lscape}
\usepackage{subcaption}

\usepackage{natbib}
\bibpunct{(}{)}{;}{a}{}{,} 

\usepackage{textgreek}
\usepackage{xargs}
\usepackage{xcolor}
\usepackage{xspace}

\let\oldtextsigma\textsigma
\renewcommand{\textsigma}{\oldtextsigma\xspace}
\let\oldAA\AA
\renewcommand{\AA}{\text{\oldAA}\xspace}
\def\w80{\ensuremath{w_{80}}\xspace}

\newcommandx{\fluxdcgs}[1][1=-20]{$\times 10^{[#1]}$~erg~s$^{-1}$~cm$^{-2}$~\AA$^{-1}$\xspace}

\newcommand{\Halpha}{\text{H\textalpha}\xspace}
\newcommand{\Hbeta}{\text{H\textbeta}\xspace}
\newcommand{\Hgamma}{\text{H\textgamma}\xspace}
\newcommand{\Hdelta}{\text{H\textdelta}\xspace}
\newcommand{\Hepsilon}{\text{H\textepsilon}\xspace}

\newcommandx{\permittedEL}[6][1=O,2=III,3=,4=,5=,6=]{\text{{#1}\,{\sc{#2}}{#3}{#4}{#5}{#6}}\xspace}
\newcommandx{\semiforbiddenEL}[6][1=O,2=III,3=,4=,5=,6=]{\text{{#1}\,{\sc{#2}}]{#3}{#4}{#5}{#6}}\xspace}
\newcommandx{\forbiddenEL}[6][1=O,2=III,3=,4=,5=,6=]{\text{[{#1}\,{\sc{#2}}]{#3}{#4}{#5}{#6}}\xspace}

\newcommandx{\HI}{\permittedEL[H][i]}
\newcommandx{\HII}{\permittedEL[H][ii]}
\newcommandx{\HeI}{\permittedEL[He][i]}
\newcommandx{\HeIL}[1][1=3889]{\permittedEL[He][i][\,\textlambda][#1]}
\newcommandx{\HeIIL}[1][1=4686]{\permittedEL[He][ii][\,\textlambda][#1]}
\newcommand{\OIII}{\forbiddenEL[O][iii]}
\newcommandx{\OIIIL}[1][1=5007]{\forbiddenEL[O][iii][\textlambda][#1]}
\newcommandx{\SIIIL}[1][1=5007]{\forbiddenEL[S][iii][\textlambda][#1]}
\newcommandx{\NeIIIL}[1][1=5007]{\forbiddenEL[Ne][iii][\textlambda][#1]}

\newcommandx{\NIL}[1]{\forbiddenEL[N][i][\textlambda][5200]}

\newcommandx{\OIL}[1][1=8446]{\permittedEL[O][i][\textlambda][#1]}
\newcommandx{\OILs}[1][1=8446]{\text{\textlambda {#1}}\xspace}

\newcommand{\OII}{\forbiddenEL[O][ii]}
\newcommandx{\OIIL}[1][1=3727]{\forbiddenEL[O][ii][\textlambda][#1]}
\newcommandx{\SIIL}[1][1=6717]{\forbiddenEL[S][ii][\textlambda][#1]}

\newcommand{\NeIII}{\forbiddenEL[Ne][iii][\textlambda][3868]}

\newcommandx{\NIIL}[1][1=6583]{\forbiddenEL[N][ii][\textlambda][#1]}

\newcommandx{\CIV}{\permittedEL[C][iv]}

\newcommandx{\CaT}{\permittedEL[Ca][ii][\textlambda][][8498,8542,8662]}
\newcommandx{\CaL}[1][1=8498]{\permittedEL[Ca][ii][\textlambda][#1]}
\newcommandx{\Ca}{\permittedEL[Ca][ii]}

\newcommandx{\Fef}{\forbiddenEL[Fe][ii]}

\newcommandx{\Fe}{\permittedEL[Fe][ii]}
\newcommandx{\FeL}{\permittedEL[Fe][ii][\textlambda]}

\newcommandx{\Feopt}{\permittedEL[Fe][ii][\textlambda][\textlambda][5190,][5320]}

\newcommandx{\CIII}{\semiforbiddenEL[C][iii]}

\newcommandx{\MgII}{\permittedEL[Mg][ii]}

\newcommand{\jwst}{\textit{JWST}\xspace}

\begin{document}

	\title{Beyond the \OIIIL[4363] auroral line: \NeIII as a direct tracer of electron temperature}
	\titlerunning{[NeIII] as a tracer of electron temperature}

	\authorrunning{B. P\'{e}rez-D\'{\i}az et al.}

    \author{Borja~P\'{e}rez-D\'{\i}az
    \inst{1}\thanks{\email{borja.perezdiaz@inaf.it}}
    \and Jos\'{e}~M.~V\'{\i}lchez\inst{2}
    \and Marco~Castellano\inst{1}
    \and Ricardo~Amor\'{\i}n\inst{2}
    \and Davide~Bevacqua\inst{1}
    \and Adriano~Fontana\inst{1}
    \and Giovanni~Gandolfi\inst{1}
    \and Antonio~Gim\'{e}nez-Alc\'{a}zar\inst{2}
    \and Laura~Pentericci\inst{1}
    \and Enrique~P\'{e}rez-Montero\inst{2}
    \and Paola~Santini\inst{1}
    \and Roberta~Tripodi\inst{1,3}}

 \institute{INAF - Osservatorio Astronomico di Roma, via Frascati 33, I-00078, Monte Porzio Catone, Italy 
         \and
         Instituto de Astrof\'{\i}sica de Andaluc\'{\i}a (IAA-CSIC), Glorieta de la Astronomía s/n, 18008 Granada, Spain 
         \and
         IFPU - Institute for Fundamental Physics of the Universe, via Beirut 2, I-34151 Trieste, Italy 
    }

	\date{Received MONTH DAY, YEAR; accepted MONTH DAY, YEAR}



\abstract
{Auroral lines enable accurate measurements of chemical abundances in ionized gaseous nebulae thanks to their sensitivity to electron temperature. However, metal-enriched systems remain a challenge, as even deep observations cannot retrieve auroral lines due to their intrinsic faintness.}
{To overcome this limitation, we present a novel approach to estimate electron temperatures in the conditions where the \OIIIL[4363] auroral line is barely detectable ($T_{e} < 11,000$ K). This approach relies on the detection of \NeIII and \OIIIL[4959,5007], which are among the brightest rest-frame optical emission lines.}
{By means of detailed photoionization models, we derive a tight relation between the O3Ne3$\equiv$\OIIIL[4959,5007]/\NeIII ratio and the electron temperature weighted in the O$^{++}$ dominated region. We test the validity of this relation in a large sample of galaxies that cover a wide range of redshifts z$\sim$0-9 and extragalactic HII regions.}
{Our results show that the O3Ne3 ratio, in combination with the O3O2 ratio (tracer of ionization), yields electron temperature estimates consistent within the uncertainties with those based on \OIIIL[4363]. The proposed relation can be used to estimate electron temperature in the cool (equivalently high-metal) regime [6,000, 13,500 K] where the emissivity of \OIIIL[4363] drops drastically.}
{}


\maketitle


\section{Introduction}
\label{intro}
The chemical composition of the interstellar medium (ISM) contains the footprints of the different processes that govern galaxy evolution. The multiphase structure of the ISM -molecular, neutral and ionized gas- requires distinct and complex modeling approaches to infer the composition of each phase\citep[e.g.][]{Cox_2005, Maiolino_2019, Schinnerer_2024}.

The ionized ISM is a proxy to understand the physical and chemical properties of the gas that feeds or has recently fed ongoing events of galaxies: young massive stars \citep[e.g.][]{Kennicutt_2012}, asymptotic giant branch phases \citep[e.g.][]{Guerrero_2013}, supernovae events \citep[e.g.][]{Hillier_2019} or feeding of supermassive black holes in Active Galactic Nuclei \citep[e.g.][]{Venturi_2018}. The ionized ISM is characterized by a moderately-high temperature (T$\sim$ 10$^{4}$ K), and is traced by prominent emission lines which carry out the bulk of the energy re-processed within the gas surrounding the ionizing source. 

The luminosity of a particular transition from a particular ion (Y$^{i}$) is given by F$_{\lambda} \equiv n(Y^{i}) \epsilon_{\lambda} (Y^{i}, n_{e}, T_{e})$, being $n(Y^{i})$ the density of the ion and $\epsilon_{\lambda} (Y^{i}, n_{e}, T_{e})$ the emissivity of the line. The latter depends on all parameters that characterize the physical conditions of the ISM (e.g. temperature $T_{e}$ and density $n_{e}$), and also on the nature of the transition (recombination or collisional excitation). The emissivities of recombination lines (RLs, e.g. Balmer lines) show little dependence on $T_{e}$ and $n_{e}$ conditions, and hence, the brightness of these lines is mainly driven by the abundance of the corresponding ions (being mainly H and He). On the contrary, as collisions are driven by the kinetic distribution of electrons, collisionally-excited lines (CELs, e.g. \OIIIL[5007], \NeIII, \NIIL[6584]) are pumped by temperature conditions, with additional dependence on density conditions due to collisional de-excitation. Hence, even for metal ions (O, N, Ne, C), which are several orders of magnitude less abundant than H or He, these CELs are bright enough to be recovered.

Inferring the ionic abundance of heavy metals from the observed metal CELs requires robust estimations of $n_{e}$ and T$_{e}$ to fully constrain the associated emissivity of the transition from available sets of atomic data \citep{pyneb, Peimbert_2017, Perez-Montero_2017}. $n_{e}$ can be estimated from doublets of emission lines (transitions from two different, yet very close in energy, excited states) due to spin-coupling to the ground level, as the ratio between their emissivities depends mostly on density. Estimating $T_{e}$ is more challenging, as it is related to the abundance of metals: the more metals are present within the ionized gas, the more coolants are available to disperse the energy, dropping the temperature. A very particular set of emission lines, associated to the transition from the second lowest to the first lowest excited states of the ions, can be used to infer $T_{e}$: the auroral lines. However, such emission lines are on the order of 1/20th (for metal-poor systems) to 1/1000th (for metal-rich systems) less bright than other CELs from the same ions. Hence, many alternative approaches, such as relying on photoionization models (e.g. \textsc{HII-CHI-Mistry} \citealp{HII-CHI_Mistry}, \textsc{NebulaBayes} \citealp{NebulaBayes}, \textsc{HOMERUN} \citealp{HOMERUN}) or using strong-line calibrations tested on galaxies for which auroral lines are retrieved \citep[e.g.][]{Curti_2017, Sanders_2024, Brazzini_2024}, have been proposed in the literature.

In this work we present a novel approach to infer electron density from \NeIII in metal-rich systems, where electron temperature is so low ($T_{e} \leq 1.1\cdot 10^{4}$ K) that auroral lines cannot be retrieved in large surveys or even with deep observations. In Sec. \ref{sec: physics} we present the physical motivation for our methodology. In Sec. \ref{sec:derivation} we discuss and provide both model-relations to derive $T_{e}$ from the \OIIIL[4959,5007]/\NeIII ratio. We review in Sec. \ref{sec:results} the performance of the new estimator. Finally, in Sec. \ref{sec:conclusions} we summarize the implications of this work.

\section{Physical motivation}\label{sec: physics}
The analysis presented in this section was done from the atomic databases listed in Table \ref{data_table} (see Appendix \ref{atomic_data} for more details).
\subsection{The case of \OIIIL[4363]}
\begin{figure}[h!]
    \centering
    \includegraphics[width=0.9\linewidth]{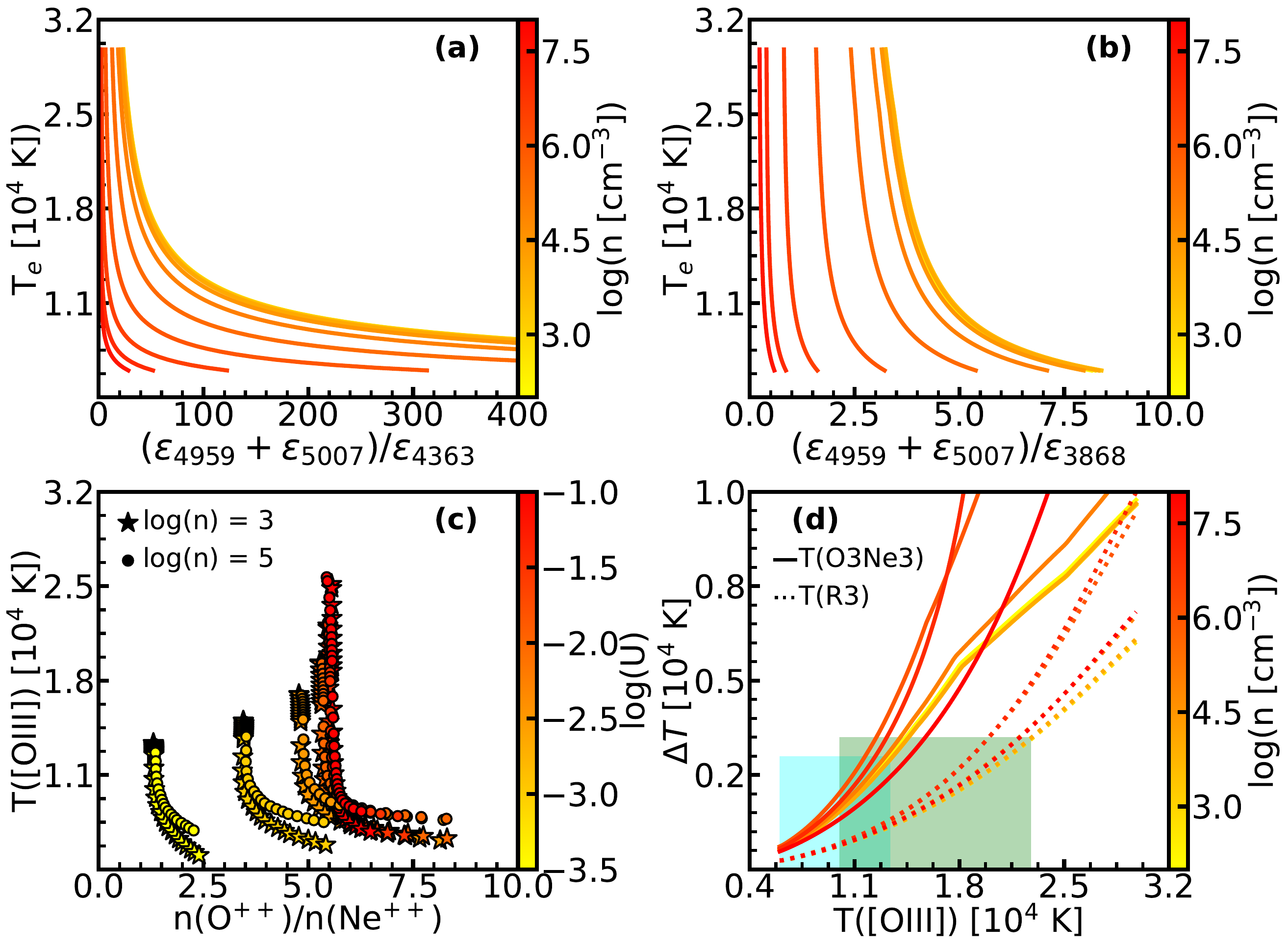}
    \caption{Theoretical behavior of the R3 and O3Ne3 emissivities. Panels (a), and (b) show emissivity ratios (x-axis) as a function of temperature (y-axis) and density (colorbar). Panel (c) shows the abundance ratio O$^{++}$/Ne$^{++}$ as a function of $T_{e}$ (y-axis) and ionization (colorbar). Panel (d) shows the associated uncertainty (for S/N $>$ 5) $\Delta T_{e}$ as a function of $T_{e}$ for O3Ne3 (continuous) and R3 (dashed) respectively. Cyan region shows the range for which the uncertainty in estimating $T_{e}$ from O3Ne3 is between [750, 3,000~K] of uncertainty. Similarly, green region denotes when the estimation from R3 is between [750, 3,600~K].}  
    \label{emissivities}
\end{figure}

The auroral line \OIIIL[4363] is the main de-excitation mechanism for O$^{++}$ to transit from $^{1}$S$_{0}$ (O$^{++}$) [5.3543 eV] to $^{1}$D$_{2}$ (O$^{++}$) [2.5146 eV]. The $^{1}$D$_{2}$ constitutes an intermediate excited state for O$^{++}$, which de-excites with a higher probability to the $^{3}$P$_{2}$ (O$^{++}$) [0.0380 eV], resulting in the prominent \OIIIL[5007] emission line, followed by the transition to the $^{3}$P$_{2}$ (O$^{++}$) [0.0140 eV] resulting in \OIIIL[4959]. In the low density regime ($n_{e} < 10^{5}$ cm$^{-3}$), electrons collide with the ground base O$^{++}$ state, and if they have enough ($\sim$ 2.5 eV) energy, they excite them to the $^{1}$D$_{2}$ level. In such conditions, more electrons can collide with the already excited ions, but they require more energy to excite them to the $^{1}$S$_{0}$ state ($\sim$ 2.8 eV). Under the assumption of a Maxwellian distribution for electrons, higher temperatures imply a higher probability not only of having $\sim$ 2.8 eV electrons, but also $\sim $ 5.3 eV (direct excitation from the ground state to $^{1}$S$_{0}$), so the \OIIIL[4363] becomes more prominent. Hence, the R3 ratio,
\begin{equation}
\label{R3} \begin{array}{rl} \mathrm{R3} = & \frac{ I\left( \mathrm{\OIIIL[4959]} \right) +  I\left( \mathrm{\OIIIL[5007]} \right)}{ I\left( \mathrm{\OIIIL[4363]} \right)} \\ = & \frac{ \epsilon_{4959} \left( T_{e}, n_{e} \right) +  \epsilon_{5007} \left(T_{e}, n_{e} \right) }{\epsilon_{4363} \left(T_{e}, n_{e} \right)} \end{array}
\end{equation}
is sensitive to $T_{e}$. The expected behavior from the characteristic emissivities as a function of $T_{e}$ and $n_{e}$ is shown in Fig. \ref{emissivities} (a).

As temperature within the ionized ISM drops (that is the case of higher metallicities, as metals efficiently cool the system), the emissivity of \OIIIL[4363] drops significantly, and difficult to detect. At the same time, for such high-metallicities, Fe emission lines such as \FeL[4352.8, 4359.3], are brighter, contaminating \OIIIL[4363] measurements. At low resolution (e.g. JWST PRISM/CLEAR observations of intermediate-$z$ galaxies), the blending between \Hgamma and \OIIIL[4363] induces more challenges in its measurement. Hence, an alternative estimation of the electron temperature becomes essential.

\subsection{The case of \NeIII}
The emission line \NeIII, originating from the transition between $^{1}$D$_{2}$ (Ne$^{++}$) [3.2038 eV] to $^{3}$P$_{2}$ (Ne$^{++}$) [ground level], has also been used as a tracer of the ionized gas at high-$z$, in the absence of \OIIIL[4959, 5007]. First of all, \NeIII is the most probable transition to de-excite $^{1}$D$_{2}$ (Ne$^{++}$), so it is the most prominent line from which Ne$^{++}$ de-excites. Secondly, the ionization energy of Ne$^{++}$ (40.96 eV) is close to that of O$^{++}$ (35.11 eV), so they mainly trace the same region within the ISM.

A crucial aspect is that the energy required to excite Ne$^{++}$ directly from the ground level is much lower than that for the \OIIIL[4363], and hence, at low temperatures, \NeIII is more prominent. Let us consider the ratio between \OIIIL[4959, 5007] and \NeIII, given by:
\begin{equation}
\label{O3Ne3} \begin{array}{rl} {\mathrm{O3Ne3}} = & \frac{ I\left( \mathrm{\OIIIL[4959]} \right) +  I\left( \mathrm{\OIIIL[5007]} \right)}{ I\left( \mathrm{\NeIII} \right) } \\ = & \frac{ n \left( \mathrm{O}^{++} \right) }{ n \left( \mathrm{Ne}^{++} \right)}\frac{ \left[ \epsilon_{4959} \left( n_{e}, T_{e} \right)  + \epsilon_{5007} \left( n_{e}, T_{e} \right) \right]}{ \epsilon_{3868} \left( n_{e}, T_{e} \right) } \end{array}
\end{equation}

As the energies for the involved excited states are lower, this ratio is not only easily observable but also sensitive to changes in temperature below a threshold from which the R3 ratio becomes asymptotic (see Fig. \ref{emissivities} b). In contrast to the \OIIIL case, the possible contaminants (\HeI, \HI lines) around \NeIII are always much fainter (< $10\%$), so the O3Ne3 ratio is robust in low resolution observations. The main caveat, compared to the R3 estimator, lies in the fact that the O3Ne3 estimator also depends on the ratio n(O$^{++}$) / n(Ne$^{++}$) which is mainly constant for high ionization parameters (log(U) > -2.0), but not under low ionization conditions due to different ionization energies (see Fig. \ref{emissivities} c).

\subsection{Critical evaluation on the tracers}
The non-linearity dependence of both R3 and O3Ne3 on $T_{e}$ requires a preliminary analysis on the conditions under which both ratios can be used for temperature estimations. For a given observed ratio (J), assuming ratios are detected at S/N $\geq$ 5, the uncertainty on $T_{e}$ amounts to $\Delta T_{e} \equiv [ T_{e} \left( 0.8\mathrm{J} \right) - T_{e} \left( 1.2\mathrm{J} \right) ]/2$.

Our results are shown in Fig. \ref{emissivities} (d) as a function of the electron temperature estimated from O3Ne3 (continous) and R3 (dashed). The lower limit in $T_{e}$ for R3 (10,500 K) is motivated due to the plateau observed in the emissivity trend from Fig. \ref{emissivities} (a), whereas for O3Ne3 we do not impose a lower limit as the plateau is not reached for the temperatures explored. The upper limits represent variations up to $\Delta T_e=$3,000 K (O3Ne3, cyan region) and $\Delta T_e=$3,600 K (R3, green region), which translates into a posterior uncertainty of ~0.3 dex in the chemical abundance derivations. The associated valid regions are represented as cyan (O3Ne3) and green (R3) regions in Fig. \ref{emissivities} (d), being the most shaded area the overlap. 

Overall, we conclude that O3Ne3 offers a unique opportunity to trace T$_{e}$ in the cool regime of 5,000 - 13,500 K, whereas R3 is still the best option to estimate T$_{e}$ beyond 13,500 K. 

\section{On the derivation of T$_{e}$ from \NeIII}\label{sec:derivation}
Due to the extra dependence of O3Ne3 on ionization, it is essential to evaluate the role of different ionizing conditions. Hence, we explore such conditions by means of photoionization models (see Appendix \ref{sec:photomodels} for more details), using as observable the O3O2$\equiv$\OIIIL[4959,5007]/\OIIL[3726,29] predictions.
\subsection{Model-based relation}
In Appendix \ref{sec:coefficients} we develop a theoretical background for the relation between O3Ne3 and $T_{e}$. In short, assuming a 5-level atom model, the expected trend is given by: 
\begin{equation}
\label{T_O3Ne3} T_{e} (\mathrm{O}^{++}) \ [10^{4}\mathrm{K}] = \frac{a}{\log\left(\mathrm{O3Ne3}\right) + b \left( \mathrm{O3O2}, n_{e} \right) } + c
\end{equation}
where $a$ and $c$ should remain essentially constant, and $b$ contains the effects of the ionizing and density conditions.

We used the grids of models from the NUVOLOSO project (see Appendix \ref{sec:photomodels} for more details) to explore the relation between the T$_{e}$ (weighted in the region dominated by O$^{++}$) and the predicted O3Ne3 ratio. 12+log(O/H) (or equivalently $T_{e}$), $n_{e}$ and log(U) are set as free parameters and we fixed the log(N/O) abundance ratio to the solar abundance, as small changes in the log(N/O) abundance does not alter our results. All models considered in this analysis are radiation-bounded. Following the methodology described in Appendix \ref{sec:coefficients}, we obtain:
\begin{equation}
\label{eq_params} 
\begin{array}{rl}
 a & = 0.17\pm0.05 \\ 
 b & = [ (0.04 \pm 0.01)\cdot\log(n_{e}) - (0.36 \pm 0.03) ]\cdot \log(\mathrm{O3O2}) \\ & + (0.08 \pm 0.02)\cdot\log(n_{e}) - (1.10 \pm 0.05) \\
c & = 0.339 \pm 0.101
\end{array}
\end{equation}

\subsection{Samples}\label{subsec:samples}
To assess the feasibility of the O3Ne3 as a tracer of $T_{e}$ we compile a sample of local galaxies (z $\lesssim$ 0.5) from SDSS DR9 \citep{Ahn_2012} and DESI DR1 \citep{DESI_DR1}, all of them being characterized by a robust (S/N $\geq$ 5) detection of \OIIIL[4363] auroral line. Additionally, we imposed that \OII[3726,3729], \OIIIL[4959,5007], \NeIII are detected with equal or higher S/N, so the tracers O3O2 (for log(U)) and R3 and O3Ne3 (for $T_{e}$) can be derived confidently. Further details on the selection criteria are provided in Appendix \ref{sec:Samples}.

We complemented our sample of local galaxies with high-$z$ galaxies (2.1 $\lesssim z \lesssim$ 9.1) observed with \jwst/NIRSpec \citep{Curti_2023, Morishita_2024, Sanders_2024}. We imposed a lower S/N $\geq$ 3 in all previously mentioned emission lines due to the limited depth in observations of high-z sources.

To explore the cool regime ($T_{e} <$ 10,000~K), we also analyzed the sample of extragalactic HII regions from massive, nearby galaxies provided by the CHAOS collaboration \citep[see][and references therein]{Rogers_2022}. After applying the same criteria as to the z$<$0.5 galaxies, we have a sample of 120 HII regions.

Overall, our test sample comprises 1,697 galaxies over a wide redshift range, and 120 extragalactic HII regions. For galaxies, the main mechanism of ionization is star formation, although we find a small group of sources (5.63$\%$) that may be affected by other mechanisms (see Appendix \ref{subsec:diagnostic}).

\section{Comparing R3 and O3Ne3 tracers}\label{sec:results}
\subsection{Evaluating the performance}
\begin{figure}[h!]
    \centering
    \includegraphics[width=0.9\linewidth]{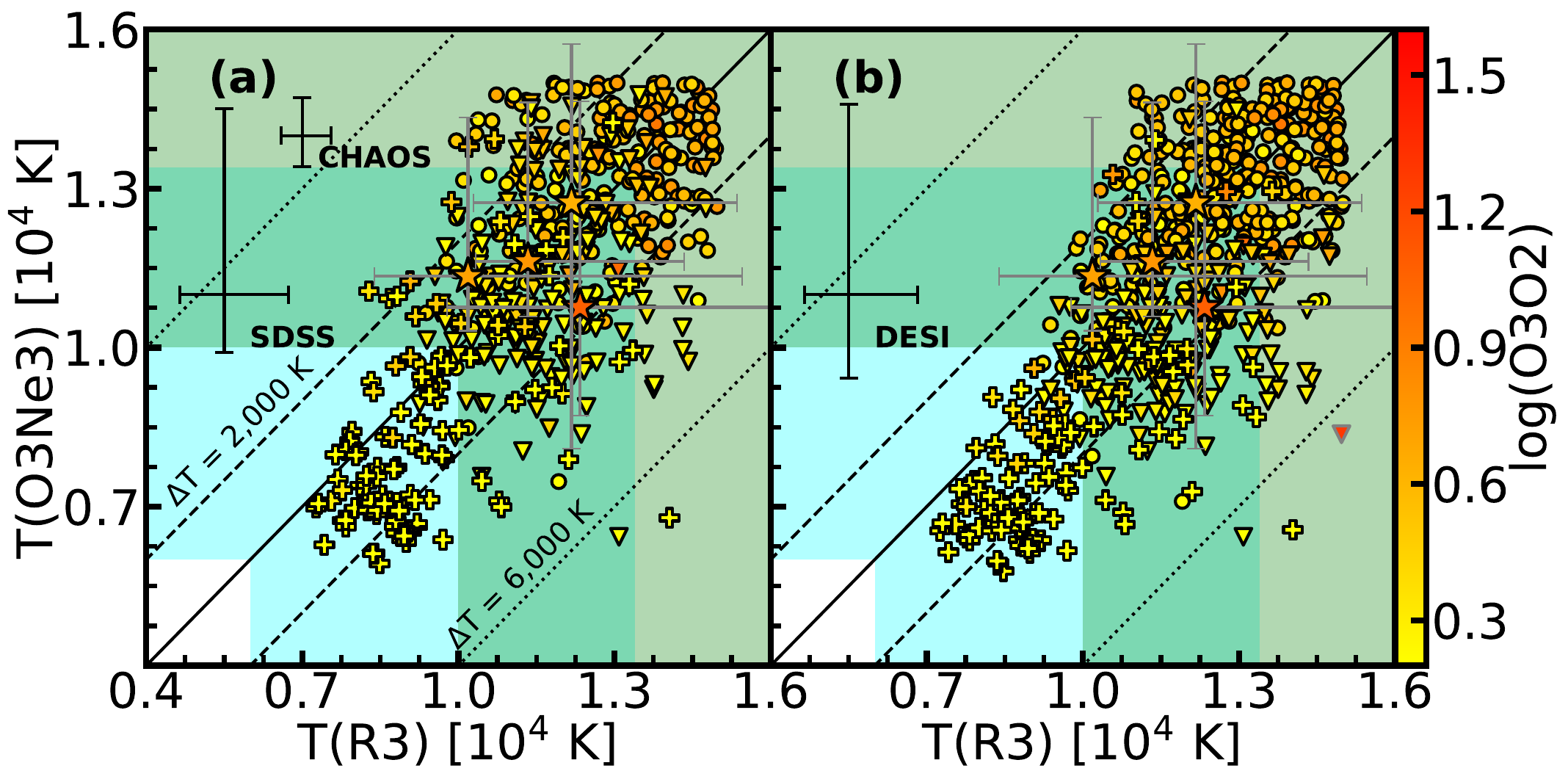}
    \caption{Temperature estimations from the R3 (x-axis) and O3Ne3 (y-axis) ratios. Circles correspond to DESI DR1, triangles to SDSS DR9, stars to galaxies from the \jwst sample and crosses to HII regions from CHAOS. Panel (a) shows the results assuming the electron density derived from the \OIIL[3726,29] doublet, while panel (b) shows the results assuming higher densities based on the O3O2 estimator. Solid black line represents the 1:1 relation, while dashed and dotted lines represent a systematic offset of 2,000~K and 6,000~K respectively. The shaded areas are the same ones as in Fig. \ref{emissivities} (d). See Fig. \ref{Results_ext} for a full version.} 
    \label{Results}
\end{figure}

As previously discussed, due to the validity range of the tracers, we will limit this study to the 6,000-15,000~K regime. In Fig. \ref{Results} we show the comparison between O3Ne3 (y-axis) and R3 derivations of $T_{e}$. We use the model-based parameters from Eq. \ref{eq_params} to convert the O3Ne3 into electron temperature. Panel (a) shows $T_{e}$ estimations with the $n_{e}$ derived from \OIIL[3727]/\OIIL[3729] ratio, and assuming $n_{e} = 1,000$ cm$^{-3}$ for those galaxies with no proper estimation of density (this includes some galaxies from SDSS and DESI and all the \jwst sample\footnote{As there is no available resolved doublet from the \jwst sample, we assumed an average value noting that this induces higher uncertainties in the temperature estimations.}). The lack of solutions when assuming the density from \OII doublet\footnote{In other words, the observed O3Ne3 ratio cannot be predicted by the models for those density and ionizing conditions.} and the incompatibility of those low densities with the high O3O2 values measured motivates the consideration of a higher density for the O$^{++}$-dominated region.

Panel (b) shows estimations assuming higher density conditions, as the diagnostics from Fig. \ref{Diagnostic} seems to favor higher densities in order to reproduce the observed O3O2 ratio. Hence, we assumed that those galaxies with log(O3O2) $>$ 1.2 are characterized by $n_{e} = 10^{4}$ cm$^{-3}$ and lower log(O3O2) are representative of $n_{e} = 1,000$ cm$^{-3}$. It must be noticed that these values are not incompatible with the lower densities estimated from \OIIL[3727,3729] lines, as there is evidence for density profiles implying higher densities for the regions responsible for the bulk of \OIII and \NeIII optical emission \citep{Castellano_2025, Hariakane_2025, Topping_2025, Moreschini_2026, Arellano-Cordova_2026}.

Fig. \ref{Results} shows that O3Ne3 provides direct estimates in agreement with the R3-based ones up to 13,500~K. Above this limit, temperatures derived from O3Ne3 systematically overestimate $T_{e}$, due to the asymptotic behavior of the emissivities shown in Fig. \ref{emissivities} (b). Converting\footnote{We note that conversions from temperature to metallicity depend on the density and, hence, the lower and upper limits represent the validity of our estimation would increase as density increases.} the $T_{e}$ range [6,000, 13,500~K] into metallicity (Z), we obtain that the O3Ne3 estimation is valid for [0.2~Z$_{\sun}$, 1.1~Z$_{\sun}$] or, in terms of 12+log(O/H), for [7.91, 8.70].

\subsection{On the origin of the scatter}
Temperature estimations from O3Ne3 in comparison to the R3-based ones show a scatter of $\sim$2,500~K for $T_{e} < 13,500$~K and $\sim$ 5,300~K for $T_{e} > 13,500$ K (see Appendix \ref{sec: relation_stats}). The scatter at high temperatures is mainly attributed to the asymptotic behavior of O3Ne3, and the presence of a large number of leaker candidates ($\sim 71\%$, see Appendix \ref{sec: relation_stats}), i.e. galaxies for which the radiation-bounded scenario explored in this work cannot be applied, as they have a high (10-20$\%$) photon escape fraction.. At low temperatures, we argue that the origin of such scatter can be found in the density conditions and/or dust depletion.

Density plays an important role in $T_{e}$ estimations. On the one hand, high density (log($n_{e}$) $\sim $ 5) conditions favor collisional de-excitation of the \OIIIL[4959,5007] transitions, making it impossible to derive robust $T_{e}$ values due to the asymptotic behavior (see Fig. \ref{emissivities} (a) and (b)). On the other hand, in the rest-frame optical spectra only doublets of low-ionized species (e.g. O$^{+}$, S$^{+}$) are available, and they might be tracing different parts of the ionized nebula. Indeed, the recent work by \citet{Arellano-Cordova_2026} points towards higher densities for the O$^{++}$ dominated region than those retrieved from the \SIIL[6717,6731] or \OIIL[3726,3729] doublets. This is highlighted in Fig. \ref{Results} when comparing the estimations for $T_{e}$ assuming the density conditions from the \OII[] doublet or higher densities, for which we obtain a tighter correlation. Associated to this problem, it is possible that some of the emission lines in galaxies come from the blending of different ionized regions with different physical conditions, but this problematic would also affect the R3 estimation.

Together with the ionization degeneracy (accounted by means of the O3O2 diagnostic), an additional problem arises from the O/Ne ratio. As Ne is also an $\alpha$-element, their nucleosynthesis origin should be similar \citep[e.g.][]{Kobayashi_2020, Esteban_2025}. Indeed, recent studies point out that the O/Ne seems to follow the solar proportions \citep{Esteban_2025}, although O depletion onto dust can reduce by 0.1 dex the ratio \citep{Mendez-Delgado_2024}, inducing a systematic\footnote{When considering instead the depletion factors provided by \citet{Jenkins_2009}, the systematic error goes up to $\sim$4,000 K} of $\sim $2,000 K, which might explain the scatter observed in Fig. \ref{Results_red} in the cool regime for objects with larger dust attenuation. 

\section{Conclusions}\label{sec:conclusions}
We present a novel approach to estimate electron temperature, essential for proper estimation of chemical abundances, based on the O3Ne3 $\equiv$ \OIIIL[4959,5007]/\NeIII ratio. We have demonstrated that the O3Ne3 ratio is an excellent approach to estimate $T_{e}$ in chemically matured (e.g. cool) regions of the ISM.

In particular, we found that Ne3O3 yields results consistent with those based on the much more challenging R3 for the temperature range $T_{e}=$ 6,000-13,500~K (e.g. 7.90 - 8.70 in terms of O abundance). Due to the asymptotic behavior of the involved emissivities, O3Ne3 yields more uncertain results for $T_{e} > 13,000$~K, precisely in the same regime where the \OIIIL[4363] is bright enough to perform a direct estimation of $T_{e}$ by means of the R3 observable.

The O3Ne3 parameter, due to its sensitivity to temperature variations in the low regime, will enable to further test the validity of strong-line calibrations for higher metallicities than those traced by direct observations of \OIIIL[4363]. Additionally, due to the brightness of the \NeIII line, such studies can be performed over statistically large samples of galaxies.

\begin{acknowledgements}
We acknowledge support from: INAF RF2024 Large Grant "UNDUST: UNveiling the Dawn of the Universe with JWST"; INAF RF2022 Mini Grant “The evolution of passive galaxies through cosmic time”; INAF RF2024 GO Grant ”Revealing the nature of bright galaxies at cosmic dawn with deep JWST spectroscopy”. JVM, RA, AGA and EPM acknowledge support from the research grant PID2022-136598NB-C32 (“Estallidos8”). EPM also acknowledges the assistance from his guide dog Rocko. 
\end{acknowledgements}

\bibliography{main}{}
\bibliographystyle{Bibtex/aa.bst}

\appendix
\section{Atomic databases}\label{atomic_data}
We used \textsc{PyNeb} \citet{pyneb} to manage the relevant atomic databases (atomic transitions, collisional coefficients and recombination rates) that are involved in the modeling of emission lines. For consistency with the photoionization models computed with \textsc{cloudy} v25 \citet{cloudy_v25}, we try to select the same databases that are used in the modeling of each emission line. Reference papers are listed in Table \ref{data_table}.

\begin{table*}[h!]
	\caption{Sources for the atomic data managed with \textsc{pyneb} \citet{pyneb}.}
	\label{data_table}
	\centering
	\begin{tabular}{c | c  c  } 
		\hline\hline
		\textbf{Ion} & \boldmath{$A_{ij}$} & \boldmath{$\Omega_{ij}$} \\ 
		\textbf{(1)} & \textbf{(2)} & \textbf{(3)} \\ \hline
		O$^{+}$ & \citet{Kisielius_2009} & \citet{FFT_2004} \\ \hline
        O$^{++}$ & \citet{SZ_2000}, & \citet{SSB_2014}  \\
         & \citet{FFT_2004} & \\ \hline
        Ne$^{++}$ & \citet{GMZ_1997} & \citet{ML_2011}  \\  
	\end{tabular}
	\tablefoot{Column (1) shows the considered ion. Column (2) gives the reference for the data containing the atomic transitions and their Einstein coefficients. Column (3) gives the reference for the collisional strengths.}
\end{table*}

\section{Photoionization models}\label{sec:photomodels}
To better explore the physical and ionizing conditions in the ionized gas that surrounds clusters of star formation, we rely on photoionization models to predict and interpret such conditions. In particular, we make use of the \textit{Nebular UltraViolet and Optical emission Lines to Optimize Studies of the iOnized gas} (NUVOLOSO) grid of photoionization models (P\'{e}rez-D\'{\i}az et al. in prep.).

In particular, we used the following grids of photoionization models:
\begin{itemize}
\item The chemical composition of the ionized nebula is governed by the 12+log(O/H) abundance, which varies in the range [6.55, 8.95] in steps of 0.1 dex. The rest of heavy elements are scaled following the solar proportion with O. Only N, which is independently varied exploring log(N/O) = [-2.0, -1.5, -1.0, -0.5, 0.0, 0.5, 1.0] and C, which is assumed to follow log(C/N) = 0.6.
\item The ionizing source is assumed to be a burst of star formation, model by means of BPASS v2.3 binary models, for which the stellar metallicity is assumed to be the same as the gas-phase, and we assume that the burst of star formation has occurred 1 Myr ago. 
\item The density conditions in the cloud are varied by exploring log(n) = [2, 3, 4, 5, 6, 7, 8] cm$^{-3}$, adopting a density-constant profile. The ionization parameter, which regulates the number of ionizing photons (e.g. the amplitude for the SED), changes in the range [-3.5, -1.0] in steps of 0.5 dex.
\item Three different scenarios for dust are included: i) the Milky Way's dust-to-gas ratio; ii) twice the MIlky Way's dust-to-gas ratio; and, iii) no dust.
\end{itemize}
In total, per density and dust content scenario, we have a grid of 2,352 models, which adds up to 49,392 models. Per model, we retrieved the information of all emission lines, with independence on how faint they are, along with the resolved ionized structure, density and temperature profiles, and the cumulative distribution of the brightest emission lines.

\section{Physical motivation for the $T_{e}$-O3Ne3 explicit formula}\label{sec:coefficients}
\subsection{The 3- and 5-level atom approach}
The physics already introduced in Sec. \ref{sec: physics} allows us to provide a physical motivation for the direct relation between O3Ne3 and $T_{e}$. Let us now consider\footnote{For the subsequent  analysis with the 3-level atom we omit the \OIIIL[4959] transition.} only $\widehat{\mathrm{O3Ne3}} \equiv$ \OIIIL[5007]/\NeIII. Assuming a 3-level atom, as the O3Ne3 is built uppon transitions from the lowest excited state to the ground level, then the explicit emissivity is obtained by following the set of equations given by \citep[e.g.][]{pyneb}:
\begin{equation}
\label{eq_balance} \sum_{j\neq i} n_{e}f_{j}Q_{ji} + \sum_{j > i} f_{j}A_{ji} = \sum_{j\neq i}n_{e}f_{i}Q_{ij} + \sum_{j < i} f_{i} A_{ij} \  \lbrace i = 1,2,3 \rbrace
\end{equation}
\begin{equation}
\label{eq_lig} \sum_{i=1}^{3} f_{i} = 1
\end{equation}
\begin{equation}
\label{eq_emissivity} \epsilon_{ij} = f_{i} A_{ij} E_{ji}
\end{equation}
where Eq. \ref{eq_balance} represents the balance between collisional ex-citations, de-excitations and spontaneous transitions, Eq. \ref{eq_lig} imposes a boundary for the values and Eq. \ref{eq_emissivity} gives the expected emissivity for the line transition. From the knowledge on the atomic transitions ($A_{ij}$) and collisional rates ($Q_{ij}$), from the tabulated values in Table \ref{atomic_data}, the fraction of ions in the state $i$ ($f_{i}$) can be obtained used for the computation of the emissivity of the line ($\epsilon_{ij}$.

As both \OIIIL[5007] and \NeIII represent the transition from $i=2$ to $i=1$, we must find the explicit form for $f_{1}$ from Eq. \ref{eq_balance} and \ref{eq_lig}. It can be demonstrated that is given by:
\begin{equation}
f_{1} = \frac{-A_{20} + n_{e}\left(Q_{02} - Q_{20}\right)}{A_{10} - A_{20} + A_{21} + n_{e} \left( -Q_{01} + Q_{02} + Q_{10} - Q_{12} - Q_{20} + Q_{21} \right)}
\end{equation}
In order to derive an approximation for the dependence on both $n_{e}$ and $T_{e}$, we can assume that $A_{ij}$ barely depends on $T_{e}$ and that $Q_{ij}$ shows the known dependence:
\begin{equation}
Q_{ij}  \propto \frac{\Omega_{ij} \left( T_{e} \right) }{T_{e}^{1/2}} \\
Q_{ji} \propto Q_{ij} \exp \left( -\frac{ \Delta E_{ij} }{k_{B}T_{e}} \right)
\end{equation}

Hence, the $\mathrm{\widehat{O3Ne3}}$ will be given by:
\begin{equation}
\widehat{\mathrm{O3Ne3}} = \frac{n \left( O^{++} \right) }{n \left( Ne^{++} \right) }\frac{f_{1} \left( O^{++} \right) A_{21} \left( O^{++} \right) E_{21} \left( O^{++} \right) }{f_{1} \left( Ne^{++} \right) A_{21} \left( Ne^{++} \right) E_{21} \left( Ne^{++} \right)}
\end{equation}
which, by taking the logarithm and expanding the expression to the first order in $n_{e}$ and $\exp \left( 1/T_{e} \right)$ yields the following approximation:
\begin{equation}
\log \left( \widehat{\mathrm{O3Ne3}} \right) \approx \tilde{b} \left( O^{++}, Ne^{++}, n_{e} \right) + \frac{\tilde{a}}{T_{e}} + O^{2} \left( T_{e}, n_{e} \right)
\end{equation}
where $\tilde{a}$ is roughly constant and $\tilde{b}$ encompasses all dependence on ionization $U$ and $n_{e}$.

To perform a more robust calculation, we use \textsc{pyneb} \citet{pyneb}, assuming a 5-level atom for both $O^{++}$ and $Ne^{++}$. We then fit the trend between $T_{e}$ and $(\epsilon_{5007}+\epsilon_{4959})/\epsilon_{3868}$ following the expression\footnote{Notice that as we are just focusing on emissivities, the $\tilde{b}$ parameter no longer show a dependence on the ionic densities. We also add an extra parameter $\tilde{c}$ to allow flexibility in the fit.}:
\begin{equation}
\label{T_emiss} T_{e} \left[ 10^{4} \ \mathrm{K} \right] = \frac{\tilde{a}}{\tilde{b} \left( n_{e} \right) + \log \frac{\epsilon_{5007}+\epsilon_{4959}}{\epsilon_{3868}} } + \tilde{c} 
\end{equation}

\begin{figure}[h!]
    \centering
    \includegraphics[width=0.95\linewidth]{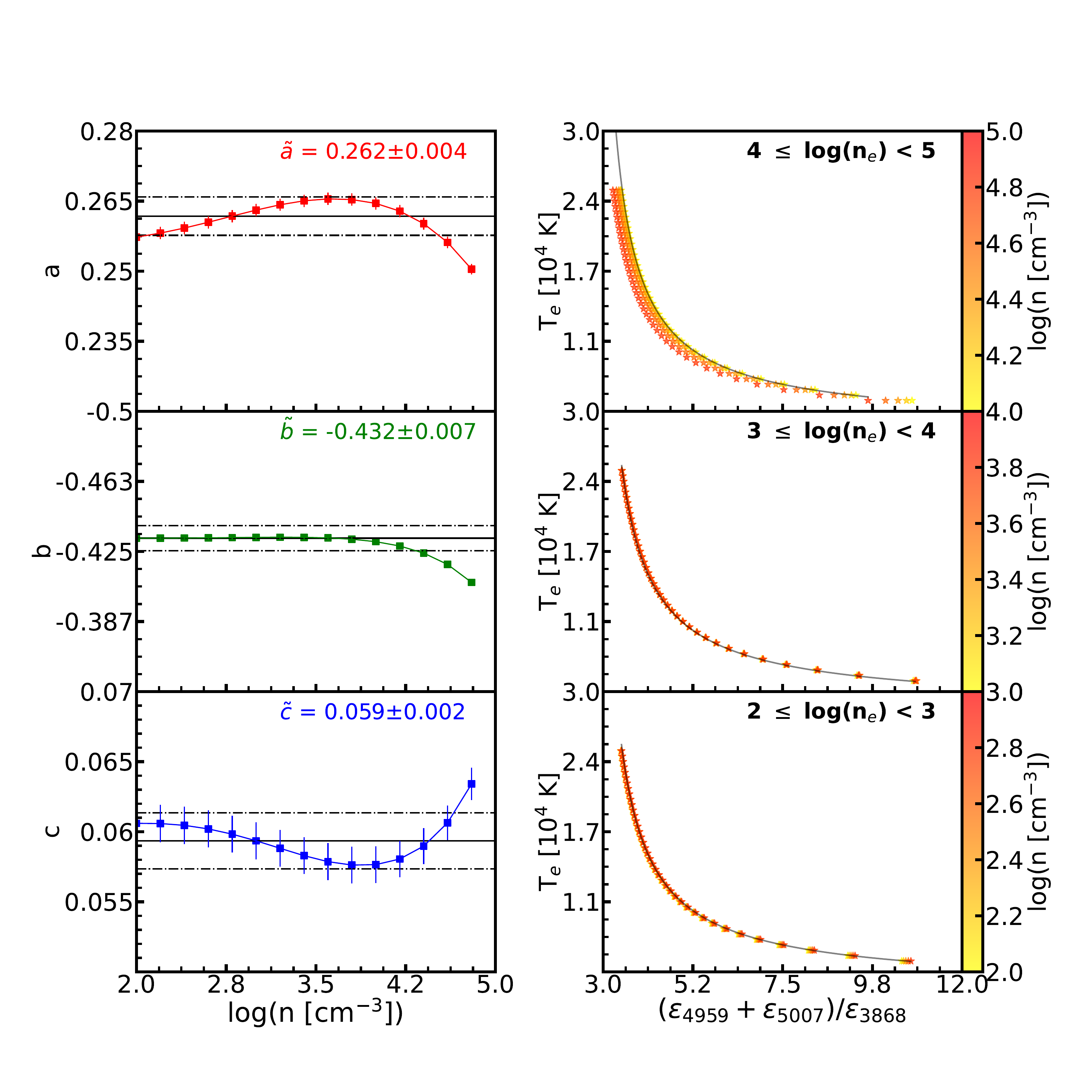}
    \caption{The global trend between $T_{e}$ and the emissivities of the O3Ne3 ratio. \textit{First column}: Best fit values for Eq. \ref{T_emiss} as a function of density. Solid and dashed lines represent the median value and its standard deviation. \textit{Second column}: $T_{e}$ as a function of the emissivity ratio for different regimes of density. Black line represents the predictions from Eq. \ref{T_emiss} assuming the best-fit parameters.}  
    \label{Fit_models_theor}
\end{figure}

As shown in Fig. \ref{Fit_models_theor}, the estimations using the \textsc{scipy} subroutine \textit{curve fit} \citep{Scipy} for $\tilde{a}$, $\tilde{b}$ and $\tilde{c}$ do not show a strong dependence on density, and if we restrict ourselves to the regime, $n_{e} \leq 10^{4}$ cm$^{-3}$ then we can safely assume that they remain constant within the uncertainties.

\subsection{Fitting the trend to photoionization model predictions}
The 5-level atom model already provides a tight relation between $T_{e}$ and the emissivity of the lines involved in the O3Ne3 ratio. As photoionization models predict the luminosity of the line, then we must introduced the explicit form for the lines. From Eq. \ref{T_emiss}, we can adopt:
\begin{equation}
\label{T_rat1} T_{e} \left[ 10^{4} \ \mathrm{K} \right] = \frac{\tilde{a} }{\tilde{b} + \log \left( \frac{n \left( Ne^{++} \right) }{n \left( O^{++} \right)} \mathrm{O3Ne3} \right) } + \tilde{c} 
\end{equation}
where we omit the dependence of $b$ on density as previously discussed. As the ratio between $n \left( Ne^{++} \right)$ and $n \left( O^{++} \right)$ can be expressed in terms of the total Ne/O abundance and the corresponding ionic fractions, then we get:
\begin{equation}
\label{T_rat2} T_{e} \left[ 10^{4} \ \mathrm{K} \right] = \frac{\tilde{a} }{\tilde{b} + \log \left( \frac{n \left( Ne \right) }{n \left( O \right) } \right) +  \log \left( \frac{f \left( Ne^{++} \right) }{f \left( O^{++} \right)} \right) + \log \left( \mathrm{O3Ne3} \right) } + \tilde{c} 
\end{equation}
where $f \left( X^{++} \right) \equiv n \left( X^{++} \right) / n \left( X \right) $ represents the ionic fraction of the element $X$. As shown in Fig. \ref{emissivities} (c), the ionic ratio shows a clear dependence on both $log U$ and $n_{e}$, which motivates the fit from Eq. \ref{T_O3Ne3}. From an observational perspective, the dependence on $log U$ can be traced by the O3O2 ratio.

\begin{figure}[h!]
    \centering
    \includegraphics[width=0.95\linewidth]{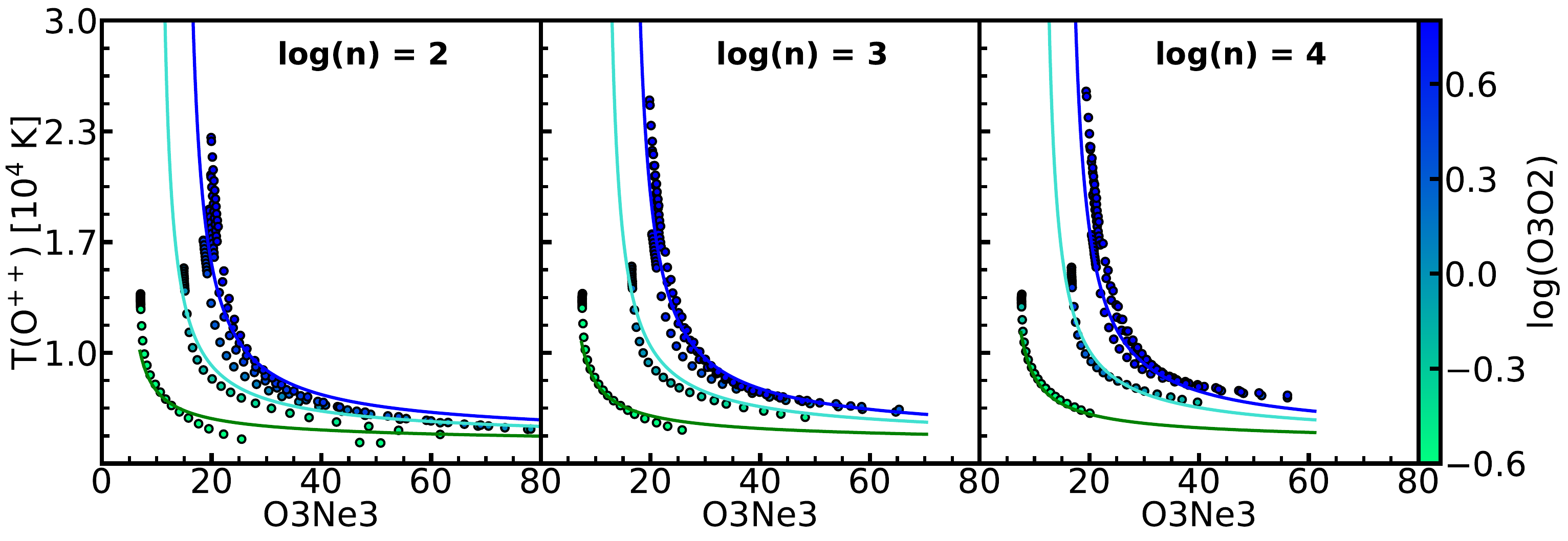}
    \caption{The predicted $T_{e} \left( O^{++} \right) $ and O3Ne3 ratios from our grid of photoionization models for different densities, color-based by the predicted O3O2. Solid lines show the fit from Eq. \ref{T_O3Ne3} with the parameters from Eq. \ref{eq_params}, assuming as representative value of O3O2 the median value for different percentiles.} 
    \label{Fit_photomodels}
\end{figure}

Starting from the initial guess that provides the 5-level atom model (see previous subsection), we fit Eq. \ref{T_O3Ne3} in three iterative steps in our grids of photoionization models. First we perform a fit allowing variations in $a$ and $c$ due to possible different treatments of the atomic databases between \textsc{Cloudy} (model predictions) and \textsc{pyneb} (5-level atom model), density is fixed at different values and five percentiles of O3O2 are explored to see the trend in $b$. Second, we fit a linear relation between O3O2 and $b$, as the statistics $R^{2} = 0.98$ point to such trend. Finally, given all the density values explored, we fit the slopes and intercepts also as linear functions of $n_{e}$ ($R^{2} = 0.95-0.99$). The fits are shown in Fig. \ref{Fit_photomodels}.

A robust check on consistency can be found on the derived form for the $b \left( n_{e}, \mathrm{O3O2} \right)$ parameter. Besides the dependence on density and ionization (O3O2) our formalism predicts a constant part which should be given by $-0.43 + \log \left( n \left( Ne \right) /n \left( O \right) \right)$. Considering the solar abundances \citep{Asplund_2009}, then this will yield $~ -1.19$, which is in agreement with the $-1.10\pm 0.05$ value derived directly from the fit to the models.

\section{Samples of galaxies and HII regions}\label{sec:Samples}
We compile a sample of galaxies that present robust measurements of the auroral line \OIIIL[4363]. As the aim of this work is to provide a robust tool to estimate electron temperatures for the analysis of the ionized gas-phase ISM, we built a heterogeneous sample of galaxies from the local Universe, observed within the Sloan Digital Sky Survey \citep[SDSS,][]{Ahn_2012}, up to $z \sim 9$, observed with \jwst. To explore the high-metal regime, we compile an additional sample of extragalactic HII regions from nearby, massive galaxies.

\subsection{SDSS Sample}\label{subsec:SDSS_sample}
From the SDSS DR9 GalSpec catalog \citep{Ahn_2012}, we selected those objects that: 1) their spectra is classified as "GALAXY"; 2) they were classified as star-forming galaxies (SFGs) according to the diagnostic diagrams\footnote{We imposed that galaxies must be classified as SFG in all four diagnostics: the BPT diagrams \citep{kewley_2006} and WHAN diagram \citep{Cid-Fernandes_2010}.}; and, 3) all emission lines were retrieved with high signal-to-noise ratio (S/N $>$ 5).

In total, we selected 257 star-forming galaxies from SDSS DR9. The redshift distribution in the SDSS sample ranges from $z\sim0$ to $z\sim 0.3$, with a median value of $z = 0.013$.

\subsection{DESI Sample}\label{subsec:DESI_sample}
From the DESI DR1 \citep{DESI_DR1}, we used the galaxy emission line catalog from \citet[][in prep.]{Zou_2024}. We apply the same filters as to the SDSS: all objects show galaxy-like spectra; the diagnostic diagrams on the emission line fluxes imply star-formation as dominant mechanism; and, all lines are retrieved with S/N > 5.

The DESI sample is composed by 1,423 galaxies, ranging from $z \sim 0$ up to $z \sim 0.5$. The median redshift for this sample is $z = 0.05$.

\subsection{JWST samples}\label{subsec:JWST_samples}
Together with the sample of local galaxies from SDSS and DESI, we compile samples of galaxies from the literature observed with \jwst. This sample of galaxies expands over a wider range of redshifts, all of them beyond z $\sim$ 2.

From \citet{Curti_2023}, we retrieved three galaxies for which we retrieved the \OIIL[3726,3729] doublet, \NeIII, \OIIIL[4363], \OIIIL[4959] and \OIIIL[5007] CELs and the Balmer lines from \Hepsilon to \Hbeta. From \citet{Sanders_2024}, we retrieved 12 galaxies with the same set of emission as to \citet{Curti_2023} sample. Finally, from \citet{Morishita_2024} we selected two galaxies, for which \OIIL[3726,3729] doublet, \NeIII, \OIIIL[4363], \OIIIL[4959, 5007]\footnote{Notice that contrary to the other samples, \citet{Morishita_2024} do not provide separate estimations for the \OIII doublet. However, given the tight relation between them due to the involved atomic transitions (\OIIIL[5007] $\approx$ 2.98*\OIIIL[4959]), we can operate with the doublet.} CELs and the Balmer lines from \Hdelta to \Hbeta are available. In summary, we compile a sample of 17 galaxies observed with \jwst, ranging in redshifts from 2.1 to 9.1.

\subsection{CHAOS HII regions}
In order to get a reliable sample of objects with deep spectroscopic observations that allow for the detection of \OIIIL[4363] in metal-rich extragalactic objects, we compile an additional sample of extragalactic HII regions from the CHAOS collaboration \citep[see][and references therein]{Rogers_2022}.

Due to the deep-, high-quality spectroscopic observations, we have also imposed that all emission lines are detected with S/N $>5$. In particular, we have selected 48 HII regions from NGC5457, 28 from M33, 18 from NGC2403, 17 from NGC628, seven from NGC3814 and two from M51. In total, our sample of extragalactic HII regions is composed by 120 objects.

\subsection{Reddening correction}\label{subsec: Reddening}
We correct for reddening attenuation by assuming the extinction curve from \citet{Cardelli_1989} and the ratio of total-to-selective extinction $R_{V} = 3.1$. For each galaxy, we select Balmer ratios that expand over the wider range of wavelength, that is \Halpha /\Hbeta for SSDS and DESI samples and \Hbeta/\Hdelta for the majority of the JWST sample. Theoretical ratios are computed assuming Case B recombination, an average electron temperature of $T_{e} = 15,000$ K and electron densities of $n_{e} = 1,000$ cm$^{-3}$, which represent the average conditions in our sample of galaxies.

For the CHAOS sample, we only retrieved the reddening-corrected emission lines and, hence, we rely on their measurements on dust attenuation. It must be noted that the authors assumed the same attenuation curve as in this work, but slightly different conditions, which are more representative of the low temperatures reported in their sample.

We show in Fig. \ref{Results_red} the comparison between electron temperatures from R3 and O3Ne3, based on their attenuation magnitude $A_{V}$. Although the O3Ne3 ratio is more affected by reddening than R3, our results show that discrepancies between both estimations are not driven by dust-attenuation effects.

\subsection{Diagnostic on the ionizing mechanism}\label{subsec:diagnostic}
Due to the lack of rest-frame red-optical coverage for most of the \jwst sample, classical diagnostics \citep{kewley_2006, Cid-Fernandes_2010} involving \Halpha and low-ionized species (e.g. \NIIL[6584], \SIIL[6717,6731]) cannot be performed. Alternatively, we can rely on the auroral line \OIIIL[4363] and the other bright nebular O lines to perform similar diagnostics.

\begin{figure}[h!]
    \centering
    \includegraphics[width=0.9\linewidth]{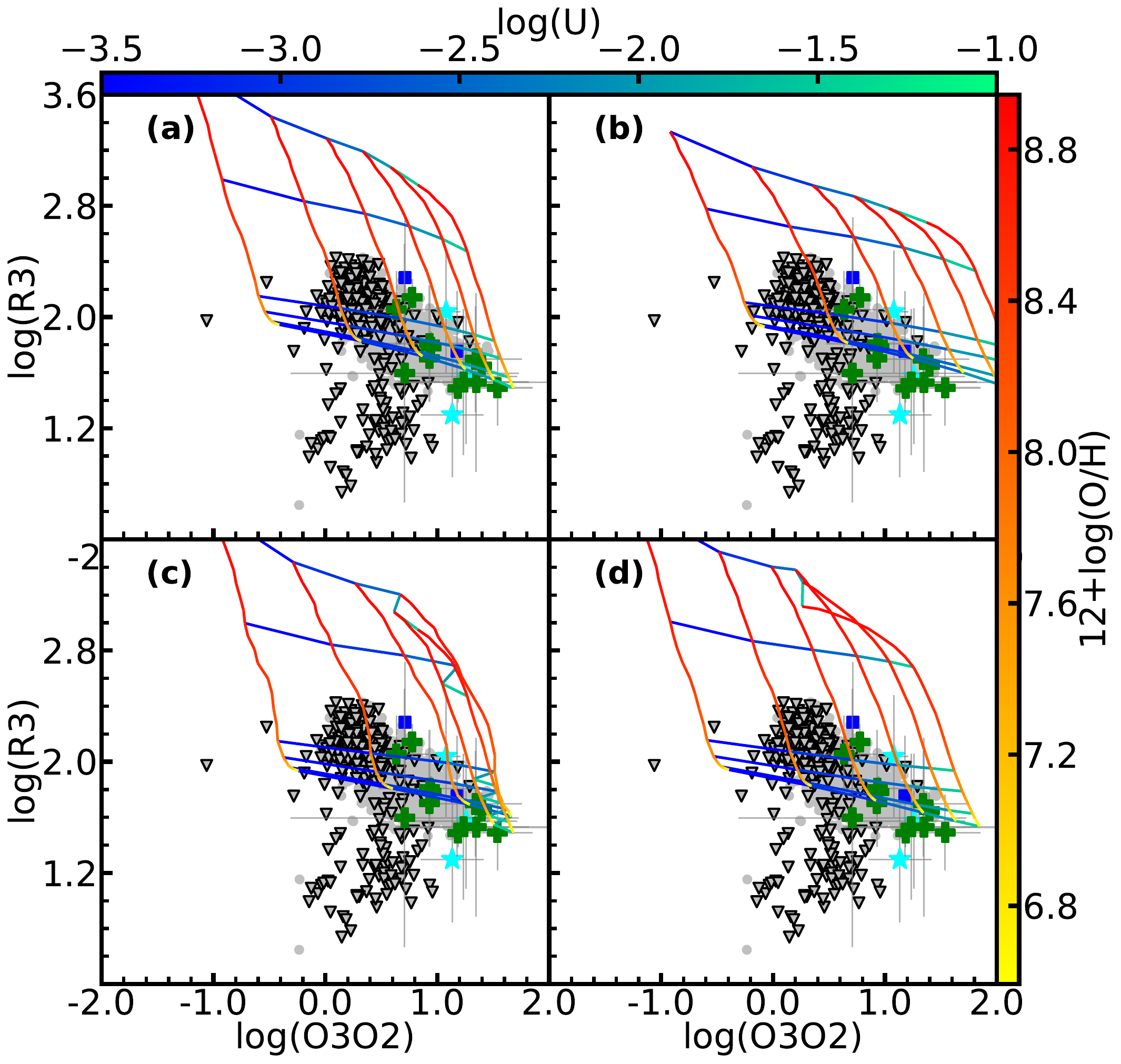}
    \caption{Diagnostic diagram R3 vs O3O2 for our sample of galaxies. Grey dots represent galaxies from DESI DR1 \citep{DESI_DR1}, triangles galaxies from SDSS DR9 \citep{Ahn_2012}, cyan stars galaxies from \citet{Curti_2023}, blue squares galaxies from \citet{Morishita_2024} and green crosses galaxies from \citet{Sanders_2024}. Magenta hexagons represent extragalactic HII regions from the CHAOS sample \citet{Rogers_2022}. Different grids of photoionization models are overlaid following the scales of the colorbars: (a) shows the grid of photoionization models for $n = 1,000$ cm$^{-3}$, (b) for $n = 10^{4}$ cm$^{-3}$, (c) shows the predictions for a matter-bounded conditions assuming that $\sim 50\%$ of \Hbeta photons escape, and (d) shows the grid of dusty-free models.} 
    \label{Diagnostic}
\end{figure}

Indeed, the ionizing conditions can be traced by means of the O3O2 ratio, whereas temperature is traced by the R3 parameter. Hence, in the R3 vs O3O2 diagram (see Fig. \ref{Diagnostic}) two trends should be expected: i) a vertical dependence on metallicity (e.g. temperature); and, ii) an horizontal dependence on ioniziation. Indeed, this is observed when superposing our grid of photoionization models on the diagram.

Due to the large coverage of our grid of photoionization models, we also observed a small sample ($\leq $ 6\%) of galaxies whose emission line ratios are not reproduced by any grid of models: neither changing the density conditions nor assuming density-bounded models. This is consistent with AGN contamination, as the \OIIIL[4363] line has been observed to be significantly brighter in AGN-dominated systems \citep[see][and references therein]{Mazzolari_2024}. Additionally, ionization due to shocks can also contribute to a brighter \OIIIL[4363] feature \citep[e.g.][]{Allen_2008}. Hence, we consider those objects with log(R3) $<$ 1.6 to be likely contaminated by other sources rather than pure star formation. 

\section{Statistical overview of the T(R3)-T(O3Ne3) relation}\label{sec: relation_stats}
Given the large number of galaxies analyzed in this work, it is necessary to review the statistics of the relation between the temperatures estimated from R3 and O3Ne3. Given the validity range for our methodology we consider two different regimes: \textit{i)} $T_{e} (\mathrm{R3}) \leq 13,500$~K (low-temperature regime); and, \textit{ii)} $T_{e} (\mathrm{R3}) > 13,500$~K (high-temperature regime).

\begin{figure}[h!]
    \centering
    \includegraphics[width=0.9\linewidth]{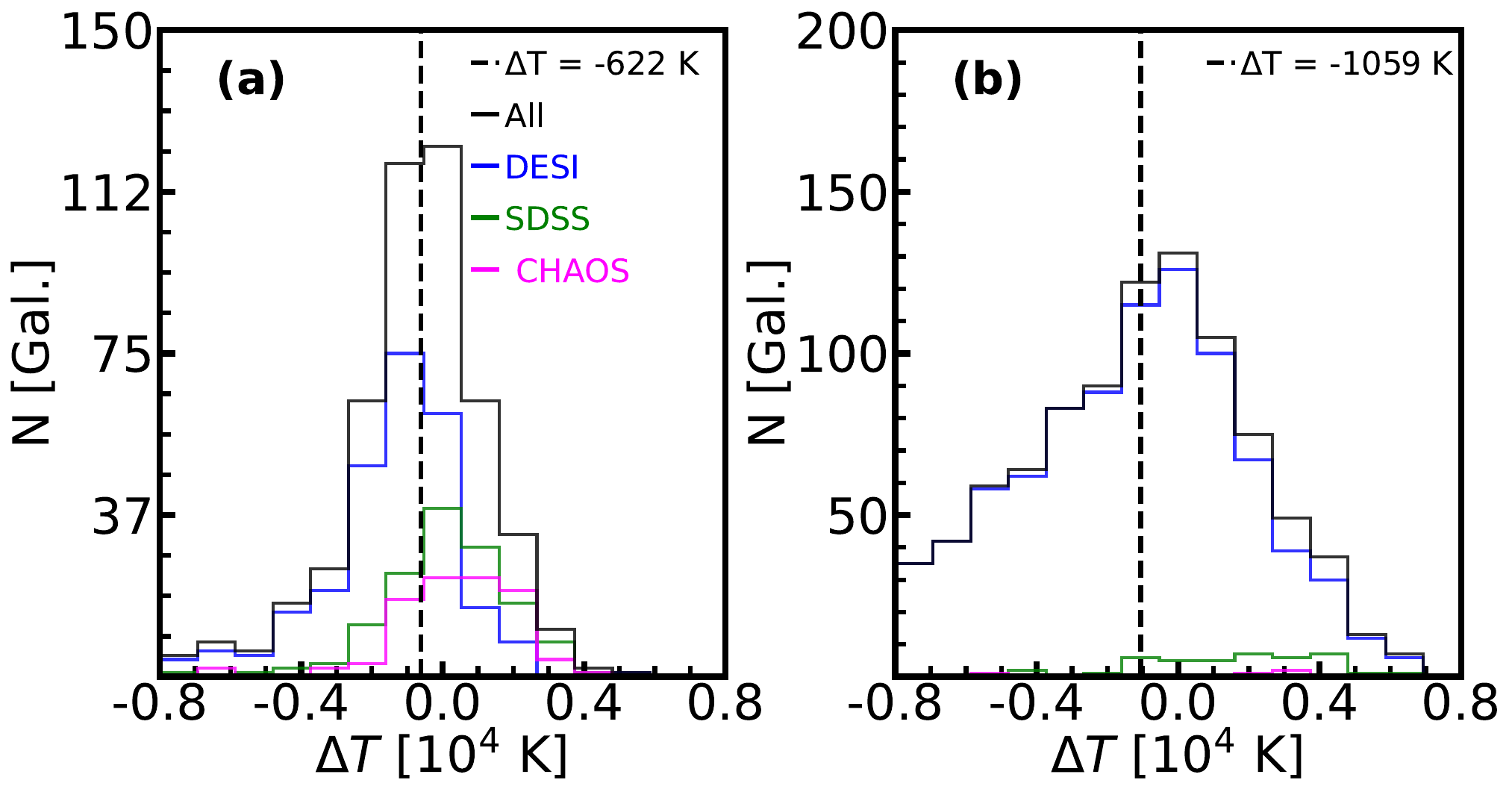}
    \caption{Histogram of the temperature difference between R3 and O3Ne3 ($\Delta T = T \left( \mathrm{R3} \right) - T \left( \mathrm{O3Ne3} \right) $). Panel (a) shows the results for those galaxies with $T \left( \mathrm{R3} \right) \leq 13,500$~K. Panel (b) shows the results for galaxies with $T \left( \mathrm{R3} \right) > 13,500$~K. Vertical dashed lines represent the median offset for each regime.}  
    \label{hist_diff}
\end{figure}

In the low-temperature regime, the Spearman test reveals a strong evidence for correlation between T(R3)-T(O3Ne3) (r $=$ 0.56, p-value $\sim 10^{-29}$). The distribution of the temperature differences ($\Delta T = T \left( \mathrm{R3} \right) - T \left( \mathrm{O3Ne3} \right) $) shown in Fig. \ref{hist_diff} (a) shows a very narrow distribution, characterized by a median offset of $\sim $- 622~K and narrow dispersion, as the root mean square error (RSME) is $\sim$2,500~K. This RSME is compatible with the propagated error from both estimations.

The results for the high-temperature regime are in agreement with the expectations from the asymptotic behavior of O3Ne3. Whereas the Spearman test also shows evidence for a correlation (r$=$0.36, p-value $\sim 10^{-30}$), both the median offset ($\sim$-1,059~K) and RSME ($\sim$ 5.300~K) are much worse than in the low-temperature regime (see also Fig. \ref{hist_diff} (b)).

It must be noted that in both regimes, the distribution is mainly governed by the DESI sample, as it provides the larger number of galaxies. From a comparative analysis of the DESI and SDSS samples, in Fig. \ref{hist_diff} we observed different trends: O3Ne3 tends to underestimate $T_{e}$ in the SDSS whereas for the DESI sample the opposite effect is observed. A similar scenario is observed if we compare the predictions for DESI and those for the HII regions from CHAOS. From Fig. \ref{Results} we can conclude that the O3O2 ratio is systematically higher for the DESI sample. Particularly, a majority ($\sim 71\%$) of the galaxies from DESI that show a stronger deviation between T(R3) and T(O3Ne3) present an O3O2 ratio above 5, which is usually indicative of leakers and that their ionizing structure is density-bounded \citep[e.g.][]{Izotov_2016}, rather than the radiation-bounded scenario considered for this work.

\section{Supplementary figures}
\begin{figure}[h!]
    \centering
    \includegraphics[width=0.95\linewidth]{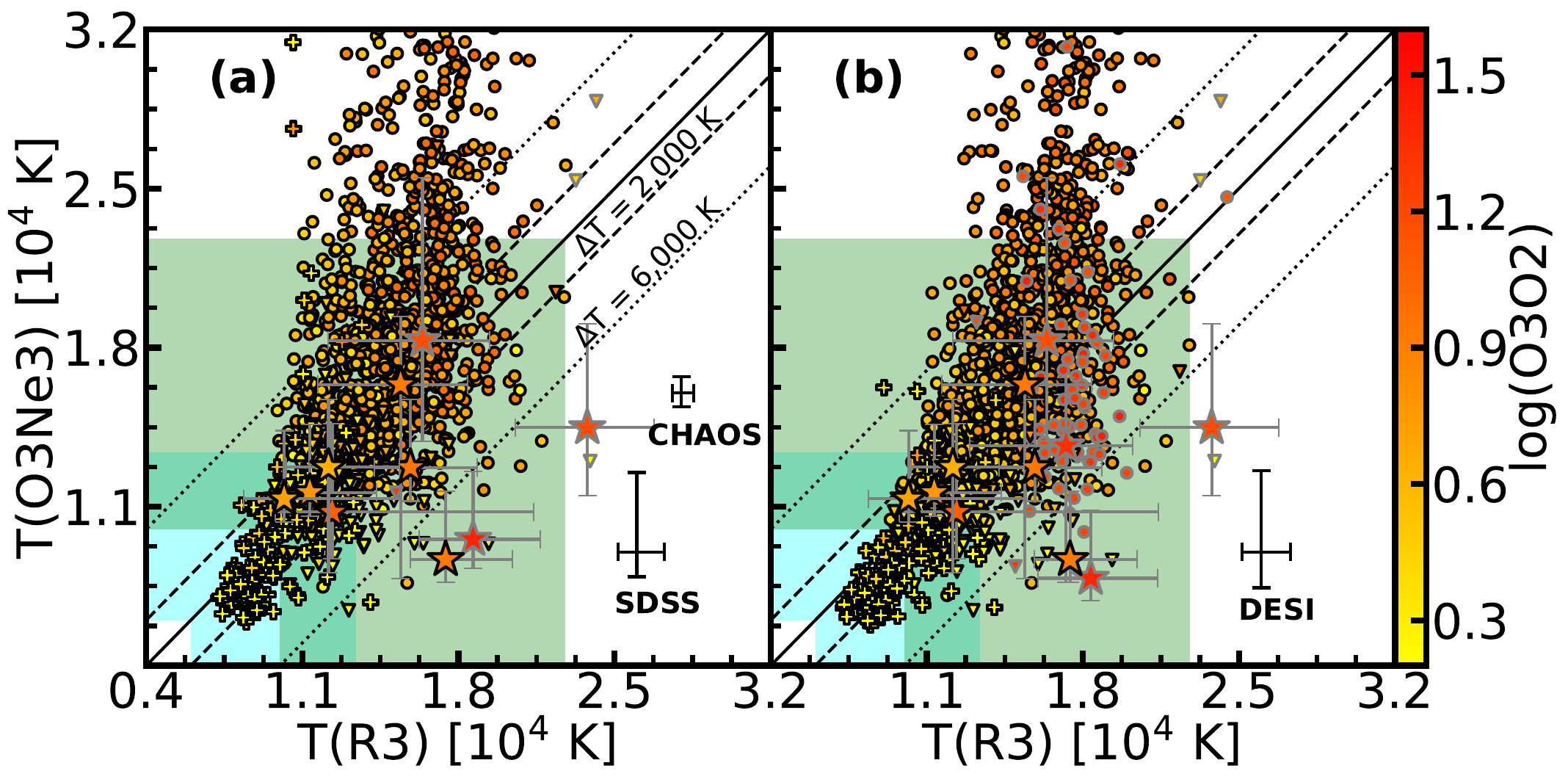}
    \caption{Extended version of Fig. \ref{Results} up to objects with $T_{e} < 30,000$~K. Black-lined elements correspond to galaxies fully covered by the grid of photoionization models, while gray-lined elements show those that do not and might be identified with potential leakers (see Appendix \ref{sec: relation_stats}).}  
    \label{Results_ext}
\end{figure}

\begin{figure}[h!]
    \centering
    \includegraphics[width=0.95\linewidth]{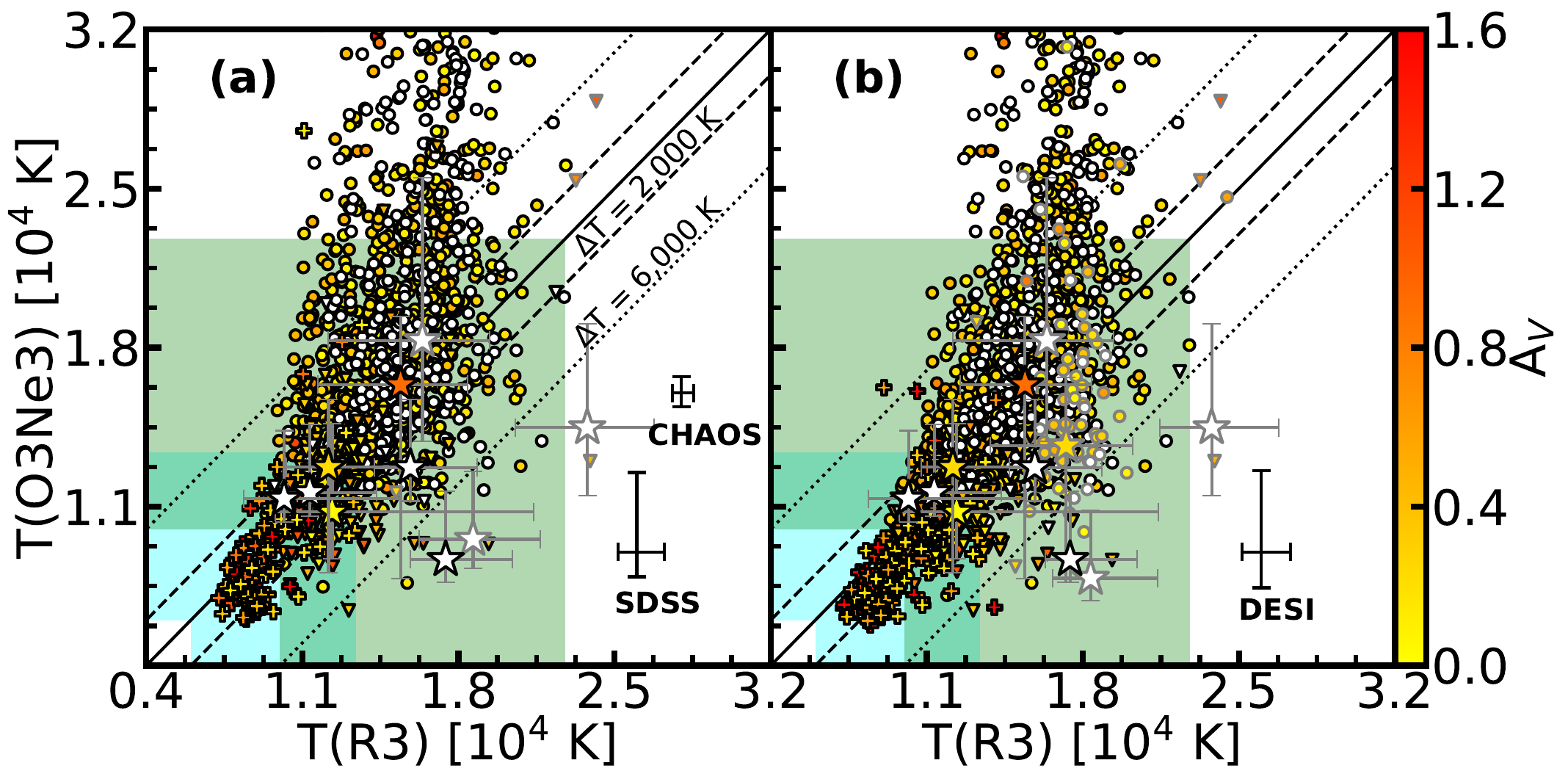}
    \caption{Same as Fig. \ref{Results_ext} but colored based on the attenuation magnitude $A_{V}$. White points correspond to galaxies with no estimation of $A_{V}$ as the observed Balmer ratio is already below the theoretical predictions.}  
    \label{Results_red}
\end{figure}

We present in this section an extended version (Fig. \ref{Results_ext}) of Fig. \ref{Results}, where we also show the behavior of galaxies characterized by higher temperatures than 15,000~K. We also present the same plot, but based on the attenuation magnitude (Fig. \ref{Results_red}), to address the impact of dust depletion effects.

\end{document}